\documentstyle[preprint,aps,epsf,eqsecnum]{revtex}
\begin{document}
\preprint{BARI-TH/95-217}
\title{$\Upsilon(4S)\to B^0 \bar{B^0} \gamma$ Background at B-Factories}
\author{P. Colangelo $^1$, G.Corcella $^{1,2}$ and G. Nardulli $^{1,2}$}
\address{
$^1$ Istituto Nazionale di Fisica Nucleare, Sezione di Bari\\
$^2$ Dipartimento di Fisica dell' Universit\'a di Bari, Bari, Italy}

\date{October 1995}
\maketitle
\begin{abstract}
{We give an estimate of the $C$-even $B^0 \bar{B^0}$ background coming from the
radiative decay $\Upsilon(4S)\to B^0\bar B^0\gamma$ at $B$-factories.
Our result $B(\Upsilon(4S)\to B^0\bar B^0\gamma) \simeq 3 \times 10^{-9}$
shows that such
background could be safely neglected in the analyses of $CP$ violating effects.
}
\end{abstract}

\setcounter{equation}{0}
\renewcommand{\theequation}{\mbox{$\arabic{equation}$}}
\newpage
The study of the process $e^+e^-\to "\gamma" \to \Upsilon(4S)\to B^0
\bar B^0$ at
the future high luminosity asymmetric $B$-factories \cite{babar,kek}
will provide us, in the next few years, with
the possibility of measuring $CP$ violation in a context
different from the kaon physics, which is the only system where it has been
observed so far. The $B$ pairs produced via decay of the $\Upsilon(4S)$
resonance offer the possibility of a clean determination of the $CP$ violating
parameter, since they are in a $C=-1$ state. For this program to be implemented
it is crucial to assess the role of the background processes where the $B^0\bar
B^0$ pair has $C=+1$, such as, for example, $B\bar B$ produced by two photon
scattering (with or without undetected $e^+e^-$ pair in the final state) or the
process $e^+e^-\to B\bar B\gamma$. Similar analyses have been performed for
$\phi$-factories \cite{dunietz,nussinov,paver} and in particular both the
$S$-wave resonant contribution to
$\phi\to K^0\bar K^0\gamma$ \cite{nussinov} and the non-resonant one
\cite{paver}
have been studied \footnote{On the whole, the predicted branching ratios
are in the range $10^{-9}-10^{-5}$ \cite{dafne}.}.

In the present work we wish to give an estimate of the
branching ratio for the process
\begin{equation}
\Upsilon(4S) \to B^0\bar B^0\gamma
\end{equation}
and to evaluate its role in
the future analysis of $CP$ violation in the $B\bar B$ system.

Differently from the case of the kaon system, we do not expect that (1) is
dominated by resonance decay into $B$ pairs, since the nearest $0^+$ resonance
is $\chi_{b0}(10235)$ which is below threshold and rather narrow
($\Gamma\simeq 0.5\ MeV$).\footnote{The full width can be estimated using the
experimental measurement of $B(\chi_{b_0}(2P)\to \Upsilon(2S)\gamma)$
\cite{pdg} and, e.g.,
the model in ref.\cite{godfried}.}
We shall study, on the contrary, the contribution
arising from the diagram in Fig.1 which is potentially important
due to the small $B^*-B$ mass difference.

In order to compute the contribution of Fig.1 we need the matrix
elements
\begin{equation}
<B^0(p_1)|J_\mu^{em}|{B^*}^0(k,\epsilon_1)>
=ig_{B^*B\gamma}\epsilon_{\mu\rho\sigma\tau}\epsilon_1^\rho p_1^\sigma k^\tau
\end{equation}
and
\begin{equation}
<{B^*}^0(k,\epsilon_1)\bar B^0(p_2)|\Upsilon(p,\eta)>
=4i\lambda_1{{M_B}\over{\sqrt{M_\Upsilon}}}\epsilon_{\mu\sigma\rho\lambda}
p_2^\mu{\epsilon_1^*}^\sigma p^\rho \eta^\lambda\
\end{equation}
($\eta$ and $\epsilon_1$ are the $\Upsilon$ and $B^*$ polarization vectors,
respectively),
where the factor $\displaystyle {{4 M_B}\over {\sqrt{M_\Upsilon}}}$
has been extracted for later convenience.

Together with (2,3) we consider the following matrix elements:
\begin{equation}
<B^0(k)\bar B^0(p_2)|\Upsilon(p,\eta)>=2\lambda_2 M_B \sqrt{M_\Upsilon}\eta^\mu
(k-p_2)_\mu
\end{equation}
\noindent and
\begin{equation}
<{B^*}^0(k,\eta_1)\bar {{B^*}^0}(p_2,\eta_2)|\Upsilon(p,\eta)>
=2\lambda_3M_{B^*}\sqrt{M_\Upsilon}\eta^\rho{\eta_1^*}^\sigma{\eta_2^*}^\lambda(
k-p_2)^\mu(g_{\mu\sigma}g_{\rho\lambda}
-g_{\mu\rho}g_{\sigma\lambda}+g_{\mu\lambda}
g_{\rho\sigma})\end{equation}
that we shall use below.

The coupling constant $g_{B^*B\gamma}$ can be written as a sum of two terms,
describing the couplings of the electromagnetic current to the heavy
($b$) and light ($q$) quark, respectively:
\begin{equation}
g_{B^*B\gamma}={{e_b}\over {\Lambda_b}}+{{e_q}\over {\Lambda_q}}
\end{equation}
($q=d$ in our case). The mass constants $\Lambda_b$ and $\Lambda_q$
behave in the $M_b\to\infty$ limit as follows: $\Lambda_b\sim M_b$ and
$\Lambda_q\sim\ const.$ They have been estimated by a number of authors
using the
heavy quark symmetries and data on $D^*\to D\gamma$ decays
\cite{amund,cho,cola},
various quark models \cite{miller,defazio} and QCD Sum Rules \cite{dosch}.
Here we take the result of Ref.\cite{cola,miller}:
\begin{eqnarray} \Lambda_b = & 5.3\ GeV\cr
\Lambda_q = & 0.51\ GeV
\end{eqnarray}
\noindent
that represent intermediate values among the various estimates.
According to the results in \cite{cola,miller}, the
theoretical uncertainty on $\Lambda_q$ (which gives rise to the dominant
contribution) should not exceed $30\%$, therefore our
conclusions should be reliable at least as order of magnitude estimates.

Let us now consider the coupling constant $\lambda_1$ in eq.(3).
Although direct experimental information on this quantity cannot be obtained
for $\Upsilon(4S)$, we can estimate it from
the knowledge of the coupling $\lambda_2$ in eq.(4) in the infinite
$M_b$ limit.
As a matter of fact, in this limit, because of the spin symmetry of the Heavy
Quark Effective
Theory arising from the decoupling of the quark spin,
the constants $\lambda_1$, $\lambda_2$ and $\lambda_3$ are related:
\begin{equation}
\lambda_1=\lambda_2=\lambda_3=\lambda \; ,\end{equation}
and their common value $\lambda$ can be estimated from the decay width of
$\Upsilon(4S)$ into $B^0\bar B^0$:
\begin{equation}
\lambda=0.7\pm 0.1\ GeV^{-3/2}
\end{equation}
where the error mainly arises from the uncertainty on the $B$ and
$\Upsilon(4S)$ masses.

The property (8) can be proved considering the $M_b\to \infty$ limit in the non
relativistic quark model formulae giving the decay widths of $\Upsilon \to BB$,
$BB^*$, $B^*B^*$ (due to phase space limitations, one has to consider $\Upsilon
(5S)$ instead of $\Upsilon(4S)$); for equal couplings ($\lambda_j=\lambda$),
the partial widths obtained using eqs. (3-5) are in the ratios
$(BB):(BB^*):(B^*B^*)=1:4:7$, as
predicted by the non relativistic quark model in the same limit
\cite{georgi,eichten}.
A different way to prove this result is as follows.
One considers an effective Lagrangian approach for the heavy mesons
consisting of one and two heavy quarks; such an approach has
been developed in \cite{casalbuoni}, where, together with the effective field
operators $H$ describing the ($B$, $B^*$) multiplet (notations as in ref.
\cite{wise}: $v^\mu$ = heavy meson velocity,
$P_\mu$ and $P_5$ annihilation operators)
\begin{equation}
H^{(b)}={{1+\not\! v}\over 2}(P^*_\mu\gamma^\mu-P_5\gamma_5)\ \ ,\ \
H^{(\bar b)}=(P^*_\mu\gamma^\mu-P_5\gamma_5)
{{1-\not\! v}\over 2}\ ,\end{equation}
also the effective field for the $Q\overline {Q}$ ($0^-,1^-$) multiplet
has been introduced
\begin{equation}
J={{1+\not\! v}\over 2}(J_\mu\gamma^\mu-J_5\gamma_5) {{1-\not\! v}\over 2}\ .
\end{equation}
By these effective fields, the Lagrangian
\begin{equation}
{\cal {L}}=\lambda \ Tr\left[\gamma^\mu\left(\partial_\mu H^{(\bar b)}JH^{(b)}
-H^{(\bar b)}J\partial_\mu H^{(b)}\right)\right]\end{equation}
can be constructed, which displays the heavy quark spin simmetry and is
completely e\-qui\-va\-lent to equations (3-5) and (8).

Using the results (6) and (9) for the coupling constants
appearing in the matrix
elements (2-5) we can now compute the decay width of the process (1) :
\begin{equation}
\Gamma={{g^2_{B^*B\gamma}\lambda^2}\over{9\pi^3}}M_B^2M_\Upsilon^3\int_{M_B}
^{{M_\Upsilon}\over 2}dE{{(E^2-M_B^2)^{3/2}(M_\Upsilon-2E)^3}\over
{(M^2_\Upsilon+M_B^2-2M_\Upsilon E)[(M^2_\Upsilon+M_B-2M_\Upsilon E-m^2_{B^*})
^2 +M_{B^*}^2\Gamma_{B^*}^2}]}\end{equation}
which gives
\begin{equation}
B(\Upsilon(4S)\to B^0\bar B^0\gamma)\simeq 3\times 10^{-9}\
\end{equation}
\noindent using $\Gamma_{B^*} \simeq 0.1 \; KeV$ as estimated in \cite{cola}.
{}From eq.(14) we evaluate that the contamination from $C$ even $B\bar B$ pairs
arising from a final state containing an undetected photon is negligible and
would not destroy the predictions for $CP$ violating effects in the process
$e^+e^-\to \Upsilon(4S)\to B^0\bar B^0$.

As a byproduct of our analysis we wish to analyze the strong decays of the
$\Upsilon(5S)$ resonance, and compute the coupling
constant $\lambda$ for the decays $\Upsilon(5S)\to BB,\ BB^*,\ B^*B^*$
defined as in the $\Upsilon(4S)$ case.

In principle, to obtain $\lambda$ from the experimental data we should also
include the decay process $\Upsilon(5S)\to B^{(*)}B^{(*)}+n\pi$ (with $n=1,2$).
We can estimate the contribution of the decay channel with one pion in the
final state by considering a diagram analogous to Fig.1, with the photon line
substituted by a pion line. The strong coupling constant $g_{B^*B\pi}$
appearing in this diagram has been theoretically studied by a number of authors
\cite{defazio,nardulli,bel}. Varying $g_{B^*B\pi}$ in the range spanned by all
these analyses we get:
\begin{equation}
{{\Gamma(\Upsilon(5S)\to B\bar B\pi)}\over {\Gamma(\Upsilon(5S)\to B\bar B)}}
=10^{-4}-10^{-3}\end{equation}
which shows that decay processes with one pion in the final state give a tiny
contribution to the full width (states with two pions are even more
suppressed because of the phase space). From this we obtain
\begin{equation}
\lambda(\Upsilon(5S))=0.07\pm 0.01\ GeV^{-3/2}\ .\end{equation}

We observe that $\lambda(\Upsilon(5S))$ is smaller than
 $\lambda(\Upsilon(4S))$ (eq.(9)), as we would expect
from a constituent quark model approach
(since the radial wave function of the bound state $\Upsilon(5S)$ has an extra
node). We also observe that,
 in computing the ratios $\Gamma(\Upsilon(5S)\to B^{(*)}B^{(*)})$,
 one should take into account the mass difference between $B$ and
$B^*$, which modifies the naive expectation $\Gamma(BB): \Gamma(BB^*): \Gamma(
B^*B^*)=1:4:7$ ; as a matter of fact we find that the widths are in the ratio
$\simeq 1: 3: 4$, which is in better agreement with the experimental data
obtained by CUSB Collaboration:
$\Gamma(BB): \Gamma(BB^*): \Gamma(B^*B^*)\simeq 1:2:4 \; \;$ \cite{cusb}.

\acknowledgments
\noindent
We thank N.Paver for discussions on the $C$-even background at a
$\phi$-factory.
We also acknowledge useful discussions on the quarkonium effective
Lagrangian with R. Casalbuoni, A. Deandrea, N. Di Bartolomeo, F. Feruglio and
R. Gatto.

\newpage
\begin{center}
\input FEYNMAN
\begin{picture}(24000,4000)
\THICKLINES
\drawline\fermion[\E\REG](0,0)[4500]
\drawarrow[\LDIR\ATTIP](\pmidx,\pmidy)
\global\advance\pmidy by -1500
\global\advance\pmidx by -1500
\put(\pmidx,\pmidy){$\Upsilon(4S)$}
\put(4500,0){\circle*{600}}
\drawline\fermion[\NE\REG](\pbackx,\pbacky)[4500]
\drawarrow[\LDIR\ATTIP](\pmidx,\pmidy)
\global\advance\pmidx by -1500
\put(\pmidx,\pmidy){$B^0$}
\drawline\fermion[\E\REG](\pfrontx,\pfronty)[4500]
\drawarrow[\LDIR\ATTIP](\pmidx,\pmidy)
\global\advance\pmidy by -1500
\put(\pmidx,\pmidy){${\bar {B^0}}^*$}
\drawline\photon[\NE\REG](\pbackx,\pbacky)[5]
\drawarrow[\LDIR\ATTIP](\pmidx,\pmidy)
\global\advance\pmidx by -1500
\put(\pmidx,\pmidy){$\gamma$}
\drawline\fermion[\SE\REG](\pfrontx,\pfronty)[4500]
\drawarrow[\LDIR\ATTIP](\pmidx,\pmidy)
\global\advance\pmidx by -1500
\global\advance\pmidy by -1500
\put(\pmidx,\pmidy){${\bar B^0}$}
\end{picture}
\vskip 2truecm
\begin{picture}(24000,4000)
\THICKLINES
\drawline\fermion[\E\REG](0,0)[4500]
\drawarrow[\LDIR\ATTIP](\pmidx,\pmidy)
\global\advance\pmidy by -1500
\global\advance\pmidx by -1500
\put(\pmidx,\pmidy){$\Upsilon(4S)$}
\put(4500,0){\circle*{600}}
\drawline\fermion[\NE\REG](\pbackx,\pbacky)[4500]
\drawarrow[\LDIR\ATTIP](\pmidx,\pmidy)
\global\advance\pmidx by -1500
\put(\pmidx,\pmidy){$\bar B^0$}
\drawline\fermion[\E\REG](\pfrontx,\pfronty)[4500]
\drawarrow[\LDIR\ATTIP](\pmidx,\pmidy)
\global\advance\pmidy by -1500
\put(\pmidx,\pmidy){${B^0}^*$}
\drawline\photon[\NE\REG](\pbackx,\pbacky)[5]
\drawarrow[\LDIR\ATTIP](\pmidx,\pmidy)
\global\advance\pmidx by -1500
\put(\pmidx,\pmidy){$\gamma$}
\drawline\fermion[\SE\REG](\pfrontx,\pfronty)[4500]
\drawarrow[\LDIR\ATTIP](\pmidx,\pmidy)
\global\advance\pmidx by -1500
\global\advance\pmidy by -1500
\put(\pmidx,\pmidy){$B^0$}
\end{picture}
\end{center}
\vskip 2truecm
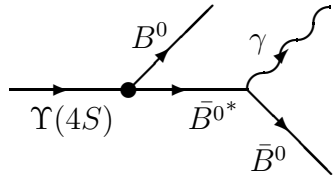
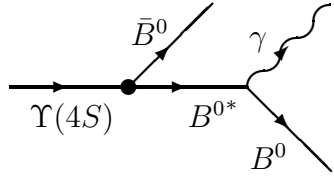
\hfil{Fig.1: Diagrams for the decay $\Upsilon(4S)\to B^0\bar B^0\gamma$.}
\end{document}